\documentclass[review]{elsarticle}
\usepackage{tabularx}
\usepackage[utf8]{inputenc}
\usepackage{lineno,hyperref}
\usepackage[top=1.65in, bottom=1.65in, left=1.35in, right=1.35in]{geometry}
\usepackage{setspace}
\usepackage{amsmath}

\usepackage{caption} 
\usepackage{algorithm}
\usepackage{algorithmicx}
\usepackage[noend]{algpseudocode}

\journal{Data \& Knowledge Engineering || arXiv preprint: 2307.15464}
\biboptions{numbers}

\makeatletter
\def\ps@pprintTitle{%
 \let\@oddhead\@empty
 \let\@evenhead\@empty
 \def\@oddfoot{}%
 \let\@evenfoot\@oddfoot}
\makeatother
\begin{document}

\begin{frontmatter}

\title{Framework to Automatically Determine the Quality of Open Data Catalogs}

\author{Jorge Martinez-Gil}
\address{Software Competence Center Hagenberg GmbH \\ Softwarepark 32a, 4232 Hagenberg, Austria \\ \url{jorge.martinez-gil@scch.at}}

\begin{abstract}
Data catalogs play a crucial role in modern data-driven organizations by facilitating the discovery, understanding, and utilization of diverse data assets. However, ensuring their quality and reliability is complex, especially in open and large-scale data environments. This paper proposes a framework to automatically determine the quality of open data catalogs, addressing the need for efficient and reliable quality assessment mechanisms. Our framework can analyze various core quality dimensions, such as accuracy, completeness, consistency, scalability, and timeliness, offer several alternatives for the assessment of compatibility and similarity across such catalogs as well as the implementation of a set of non-core quality dimensions such as provenance, readability, and licensing. The goal is to empower data-driven organizations to make informed decisions based on trustworthy and well-curated data assets. The source code that illustrates our approach can be downloaded from \url{https://www.github.com/jorge-martinez-gil/dataq/}.
\end{abstract}


\end{frontmatter}

\section{Introduction}
Efficient data management strategies have become a critical need for a wide range of organizations, leading to the use of data catalogs (DCs) as a helpful strategy to facilitate data management. As metadata repositories, DCs are essential in identifying, evaluating, and using diverse data assets. However, guaranteeing the quality of DCs presents a complex challenge, with particular importance in dynamic data ecosystems where the contents evolve rapidly, as is the case of the Open DCs (ODCs)\cite{key-abs2}. The manual assessment processes are time-consuming, resource-intensive, and even error-prone. Therefore, developing sound strategies for quality assessment has attracted much attention in recent times \cite{key-albertoni}.

In response to this challenge, this research presents a novel framework specifically designed to automate the evaluation of the quality of ODCs. The rationale for this framework lies in recognizing that ODCs should publicly supply contextual information about the cataloged data assets. Therefore, it is necessary to expose details such as data sources, owners, policies, and any other accompanying information \cite{key-Ryan}. Such contextual information should facilitate stakeholders with a good understanding of the data's origin, purpose, and limitations.

There is a certain consensus in the community that ensuring the quality of ODCs deserves further research on quality aspects \cite{key-albertoni}. For example, suppose an organization wants its ODCs to be considered a trustworthy information source. In that case, it must entrust people to confidently use appropriate data assets for their analytical and decision-making efforts \cite{key-labadie}. Contrarily, a catalog of subpar quality can result in wasted time, erroneous analysis, and reduced decision-making capabilities. Thus, developing a framework that automates the assessment of catalog quality assumes particular significance, at least for most Open Data (OD) consumers.

Our proposed framework involves multiple quality dimensions: accuracy, completeness, consistency, scalability, timeliness, compatibility, similarity, provenance, readability, and licensing. These dimensions have various aspects that determine the reliability and utility of cataloged data assets. Our framework aims to provide insights for users by adopting a practical approach to quality measurement. The idea is that stakeholders can also easily understand, use, and extend the proposed methods.

Furthermore, to demonstrate the capabilities of our framework, we illustrate our definitions with implementations using DCAT\footnote{W3C's Data Catalog vocabulary}. This vocabulary is the de facto metadata standard within the domain of OD. It represents a W3C recommendation for describing OD on the Web \cite{key-abs,key-maali}. It is based on the RDF vocabulary and enjoys wide acceptance at present \cite{key-Neumaier}. Therefore, the most significant contributions of this research work can be summarized as follows:

\begin{itemize}
	\item This research introduces a novel framework that automates the evaluation of ODC quality, effectively tackling the difficulties inherent in manual assessment procedures. The proposed framework shows potential in facilitating decision-making processes driven by data and optimizing the overall value obtained from data assets.

  \item We also delineate ODC cross-catalog dimensions and explore potential implementations for cases involving multiple catalogs. We develop the notions of compatibility and similarity to contribute to a comprehensive understanding and development of new quality metrics in ODC quality assessment.

  \item All definitions provided are mapped to existing works and ISO standards, and an exemplary implementation using the W3C's Data Catalog standard is provided. This in-depth exploration illustrates the practical applicability of the proposed framework and the theoretical principles discussed in the paper.
\end{itemize}

The subsequent sections of this work are organized as follows: Section 2 explains the existing landscape regarding the utilization of ODCs. Section 3 encloses our contribution regarding core quality dimensions, encompassing two pivotal elements: a formal definition of the vital quality concepts and a suggestion outlining the pseudocode implementation. Section 4 offers new developments regarding cross-catalog dimensions, including the compatibility and the similarity between ODCs. Section 5 introduces several non-core dimensions that are still very useful in facilitating the tasks of data stakeholders. Section 6 introduces several actual use cases and a discussion. Finally, this research work's conclusion is presented based on our findings and discusses potential lines for future research.

\section{State-of-the-art}
An ODC is a metadata repository that organizes information about the available data assets within a public audience \cite{key-Lakomaa}. It is a reference for data consumers, e.g., analysts, scientists, etc. The aim is to facilitate the discovery, understanding, and access to relevant data \cite{key-Debattista}. 

\subsection{Data catalogs and Metadata quality}
A well-designed ODC should combine the relevance of cataloged data assets with the needs and interests of its users. It should offer, at least, efficient search capabilities and clear navigation options \cite{key-Pietriga}. Usability aspects such as collaborative features can also improve the user experience \cite{key-Klime}. At the same time, an ODC should facilitate effective filtering mechanisms, allowing users to find data assets aligned with their needs.

ODCs should also adhere to data governance principles and support effective metadata management. This includes maintaining consistent metadata standards, establishing quality rules, enforcing access controls, and facilitating integration with other tools and processes. Therefore, the importance of an ODC lies in its ability to address some of the traditional challenges associated with data management. The literature provides plenty of reasons why ODCs play or should play an important role:

\begin{itemize}
	\item An ODC is a valuable tool for data discovery, providing an inventory of available data sources, datasets, etc. It helps users efficiently locate relevant data assets \cite{key-Barthelemy}.
	
	\item An ODC enhances data understanding by providing metadata about the data assets, including detailed descriptions and relationships with other datasets. This information helps users evaluate the data's structure, content, and context for analysis or decision-making purposes \cite{key-geissner}.
	
	\item An ODC plays a pivotal role in data governance by establishing standards, policies, and guidelines for presenting a consistent view of an organization's data assets \cite{key-Neumaier}. Data managers can enforce governance rules, monitor data utilization, and comply with internal and external regulations.
	
	\item Collaboration and knowledge sharing among users are promoted through an ODC. Users can annotate datasets, provide feedback, share insights, and collaborate on projects. The catalog reduces data silos, encourages sharing, and promotes a data-driven culture \cite{key-ochoa}.
	
	\item An ODC facilitates data lineage by providing information on data assets' origins, transformations, and dependencies, allowing users to trace the data journey and promote transparency. It also enables impact analysis when identifying the effects of changes to data sources  \cite{key-BNK}.
	
	\item An ODC assists in ensuring data security and compliance with privacy regulations \cite{key-dibowski}. It documents data access controls, classification labels, and sensitive elements, helping enforce security measures and monitor data access. 
\end{itemize}

However, there is still much to be investigated, as the growing number of research papers in this regard testifies \cite{key-Ehrlinger}. The aim is to achieve comprehensive solutions that enable the effective and efficient use of metadata by individuals or systems acting on their behalf.

\subsection{The DCAT vocabulary}
The DCAT vocabulary is an established standard for effectively describing ODCs on the Web. DCAT involves a set of classes, properties, and relationships that facilitate the publication, discovery, and exchange of essential information about datasets \cite{key-Nogueras}. This standardization facilitates interoperability across diverse platforms and systems, thus enabling the integration and utilization of data resources \cite{key-Lobe}.

One of the fundamental contributions of DCAT lies in defining concepts such as catalogs, datasets, distributions, and data services. This vocabulary entrusts data publishers to thoroughly describe pertinent metadata related to datasets, including but not limited to title, description, keywords, licensing information, spatial and temporal coverage, and so on. Moreover, DCAT provides robust support for establishing connections between datasets and associated resources such as data files or APIs \cite{key-Kirstein}.

The standardized representation of data resource information through DCAT simplifies exposing datasets for data publishers while facilitating OD consumers' efficient data search. Furthermore, this standardized vocabulary employs a shared vocabulary and enables data exchange and integration among diverse data platforms, data portals, and other data-centric tools.

DCAT is commonly used by organizations aiming to improve the visibility of their data assets effectively. The idea is to promote data sharing and reuse while guaranteeing interoperability within a broader data ecosystem. Therefore, adopting DCAT is pivotal in facilitating transparency and collaboration through efficient data resource management.

\subsection{Research gap}
Numerous challenges persist in quality assessment for ODCs. Various issues regarding data quality emerge, enclosing inconsistency, inaccuracy, incompleteness, duplication, outdated information, and human error \cite{key-Gupta}. These issues have a profound influence on decision-making procedures. Existing tools designed to address the problem exhibit imperfections and predominantly rely on domain-specific niches. This is primarily due to the multifaceted nature of quality assessment within data repositories. While some standard practices exist, individual approaches to ensuring data quality have historically demonstrated superior performance.

Consequently, more research is needed concerning standardized metrics to evaluate the quality of ODCs. Indeed, the absence of widely accepted metrics capable of independently assessing key aspects like accuracy, completeness, and consistency of ODCs, irrespective of the application domain, is evident. Moreover, as ODCs expand in size and complexity, evaluating their scalability potential automatically poses a formidable challenge. Consequently, research efforts must develop efficient techniques to assess large-scale data catalogs' timeliness while maintaining other pertinent aspects.

\subsection{Contribution over the state-of-the-art}
As Table \ref{tab:tab0} shows, several works \cite{key-Kubler, key-Zaveri} and international standards (ISO 19157\footnote{https://www.iso.org/standard/32575.html}, ISO 25012\footnote{https://www.iso.org/standard/35736.html}) on assessing quality have been proposed in the literature. However, the nomenclature used is not homogeneous, and the assessment methods are not formally defined, which is detrimental to a framework for a fair comparison. Our work offers, for the first time:

\begin{enumerate}
	\item A focus on the automatic quality assessment of ODCs,
	\item A formal definition for each quality dimension that significantly facilitates their understanding and matching with other metrics proposed in the literature\footnote{Please note that, unlike other approaches, this research does not consider accessibility as a quality dimension since this notion is covered in the definition of accuracy},
	\item An implementation proposal for each of the proposed quality dimensions to facilitate research progress transparency and
	\item A proposal for automatically determining compatibility and similarity between ODCs.
\end{enumerate}
 
This framework is organized around three types of quality dimensions: Core (accuracy, completeness, consistency, scalability, and timeliness), Cross-catalog (compatibility and similarity), and Non-Core (lineage and provenance, readability, and licensing). Furthermore, all these contributions orbit around the DCAT standard -on which few studies are yet- as a reference for this research work.

\begin{table}
\caption{Alignment of the quality metrics proposed in the literature concerning the present work}
\label{tab:tab0}
\begin{tabularx}{\textwidth}{|p{2.55cm}|p{2.3cm}|p{2.6cm}|p{2.65cm}|X|}
\hline
\textbf{Zaveri et al.} & \textbf{Kubler et al.} & \textbf{ISO 25012} & \textbf{ISO 19157}                     & \textbf{This work}     \\
\hline
Accuracy               & Accuracy               & Accuracy           & Correctness                            & Accuracy                  \\
Completeness           & Existence              & Completeness       & Completeness                           & Completeness              \\
Consistency            & Conformance            & Consistency        & Consistency                            & Consistency               \\
Timeliness             & -                      & Currentness        & Temp. Validity                      		& Timeliness                \\
Availability           & Retrievability         & Availability       & Non-q. Correct. 												& Scalability               \\
Accessibility          & Open Data              & Accessibility      & Non-q. Correct. 												& - \\
Possibility            & Open Data              & Portability        & Consistency                    				& Licensing                 \\
Understandability      & -                      & Understandability  & Text Quality	                   				& Readability               \\
-                      & -                      & Compliance         & -                                      & Provenance    \\
-                      & -                      & -                  & -                                      & Compatibility             \\
-                      & -                      & -                  & -                                      & Similarity \\
\hline              
\end{tabularx}
\end{table}

\section{Core Quality Dimensions in Open Data Catalogs}
At first instance, this research assumes an ODC should offer several core characteristics for evaluating the intrinsic properties of the data and, therefore, contribute to its effectiveness in facilitating discovery, understanding, and utilization. The following are those core quality dimensions: accuracy, completeness, consistency, scalability, and timeliness. Please note that all definitions assume that we refer to a well-formed ODC according to the rules of the DCAT standard. It is also assumed that we do not have access to the underlying assets cataloged, so all operations are performed on information contained only and exclusively in the catalog.

\subsection{Accuracy}
ODCs should maintain accurate information about the data assets they represent. The metadata should reflect the actual characteristics of the underlying data. Ensuring accuracy helps users avoid errors when working with the cataloged data. Therefore, in the context of an ODC, data accuracy should be associated with the correctness of the information provided about the data.

\subsubsection{Accuracy in a Data Catalog}
Accuracy is the quality dimension that aims to quantify the extent to which $\mathcal{C}$ contains correct information. For the remainder of this paper, and as the DCAT standard is based on RDF, please consider attributes as the objects of the assertions and relations as the predicates.

\subsubsection{Attribute-level Accuracy}
Attribute-level accuracy refers to the correctness of the attribute values in the ODC. Let $V(a_k)$ be the actual value associated with attribute $a_k$, and $D(a_k)$ be the corresponding value in the ODC.

The attribute-level accuracy for $a_k$ can be measured using an error function $E(a_k)$, which assesses the discrepancy between the actual value $V(a_k)$ and the stored value $D(a_k)$. Therefore, the attribute-level accuracy for an ODC can be defined as shown in Equation \ref{eq:aa}.

\begin{equation}
\text{Attribute-level Accuracy}(\mathcal{C}) = \frac{1}{m} \sum_{k=1}^{m} (1 - E(a_k))
\label{eq:aa}
\end{equation}

where $m$ is the total number of attributes in the ODC.

Please note that this definition involves the average discrepancy across all attributes, providing an overall measure of the ODC accuracy concerning its attributes.

\subsubsection{Relationship-level Accuracy}
Rel.-level accuracy refers to the correctness of relationships in the ODC. Let $R(a_k, a_l)$ be the actual relationship between attributes $a_k$ and $a_l$, and $C(a_k, a_l)$ be the corresponding relationship recorded.

The rel.-level accuracy for $a_k$ and $a_l$ can be measured using an error function $E(a_k, a_l)$, which assesses the discrepancy between the true relationship $R(a_k, a_l)$ and the actual relationship $C(a_k, a_l)$ exactly as shown in Equation \ref{eq:ra}

\begin{equation}
\text{Rel.-level Accuracy}(\mathcal{C}) = \frac{1}{\binom{m}{2}} \sum_{k=1}^{m-1} \sum_{l=k+1}^{m} (1 - E(a_k, a_l))
\label{eq:ra}
\end{equation}

where $\binom{m}{2}$ is the total number of unique attribute pairs in the ODC.

This definition again considers the average discrepancy across all pairs. The idea is to have an overall measure of the accuracy of the catalog's relationships.

\subsubsection{Overall Accuracy}
The overall accuracy of $\mathcal{C}$ combines both accuracy levels as shown in Equation \ref{eq:oa}.

\begin{equation}
\text{Accuracy}(\mathcal{C}) = \alpha \cdot \text{Attr.-level Accuracy}(\mathcal{C}) + (1-\alpha) \cdot \text{Rel.-level Accuracy}(\mathcal{C})
\label{eq:oa}
\end{equation}

where $\alpha$ is the weighting factor about the relative importance of the two types of accuracy seen above.

It is difficult to dictate the values that should ideally be represented generically. However, the absence of particularly significant attributes (attribute level), broken links (attribute level), or duplicate information (relationships level) is detrimental to the accuracy of an ODC. The following is an elaboration of this intuition.

\subsection{Completeness}
An ODC should cover an organization's available data assets, ensuring users can quickly locate and access relevant data.

\subsubsection{Completeness in a Data Catalog}
Completeness is the quality dimension that aims to quantify the extent to which the necessary information is captured in the ODC. This means that, for each object $O_i$ with attributes $A_i = {a_1, a_2, a_3, \ldots, a_m}$, the Equation \ref{eq:comp} must hold.

\begin{equation}
\forall a_k \in A_i, \exists \text{{description}}(a_k, O_i)
\label{eq:comp}
\end{equation}

where $\text{{description}}(a_k, O_i)$ represents the presence of a description for attribute $a_k$ in $O_i$. 

Furthermore, completeness also requires documenting all essential relationships among attributes. Therefore, for any pair of attributes $a_k, a_l \in A_i$, the condition expressed in Equation \ref{eq:comp2} must hold.

\begin{equation}
relationship(a_k, a_l) \Rightarrow doc.\_relationship(a_k, a_l, O_i)
\label{eq:comp2}
\end{equation}

where $relationship(a_k, a_l)$ is an actual relationship between the attributes $a_k$ and $a_l$, and $doc.\_relationship(a_k, a_l, O_i)$ is the description for the relationship between $a_k$ and $a_l$ in $O_i$. Therefore, $\mathcal{C}$ is complete if the above conditions hold for all $O_i$ in $\mathcal{C}$.

The goal is to check if a given ODC contains descriptions for all attributes, capturing their essential characteristics. It also provides documentation for relationships, facilitating the understanding of the data.

\subsection{Consistency}
Consistency is crucial for an ODC to provide reliable information and ensure internal consistency. The catalog should follow consistent naming conventions, data formatting standards, and metadata representations across all entries. However, any interpreter would detect a syntactic inconsistency of these types. Here, we refer to a semantic type of inconsistency. Consistency promotes ease of use, trust, and integration with other systems.

\subsubsection{Consistency in a Data Catalog}
Checking for inconsistencies in a given $\mathcal{O}$ of $\mathcal{C}$ involves the identification of contradictory triples $(t_i, t_j) \in T \in \mathcal{O} \in \mathcal{C}$. Two triples belonging to the same $\mathcal{O}$ are deemed contradictory if, for two intimately related triples $(t_i, t_j)$, the value of their respective objects is not possible. We can express it formally through the Equation \ref{eq:cons}.

\begin{equation}
    \text{Contradictory}(t_i, t_j) = \neg \left( \text{ValidRel.}(t_i) \land \text{ValidRel.}(t_j) \land \text{Rel.-Compatible}(t_i, t_j) \right)
    \label{eq:cons}
\end{equation}

The average (in)consistency of the ODC is the proportion of (in)consistent objects in the catalog. Some examples could be: the modification date is earlier than the creation date, unique labels are duplicated, a tag specifies that the language is English, but the text is in Spanish, or a triple indicates a file in \textit{json} format, but the \textit{url} returns a spreadsheet in proprietary format, etc.

\subsection{Scalability} 
ODCs should be scalable to accommodate growing data volumes and evolving data landscapes. In reality, ODCs are not scalable per se but rather the computational methods that manage them. However, this dimension is critical, and the goal is to build a procedure that elucidates whether it is possible to use scalable methods on the ODC. The idea is to measure the capability of the catalog to expand and evolve alongside the organization's changing data ecosystem.

\subsubsection{Scalability in a Data Catalog}
Scalability refers to efficiently handling an increasing volume of data and changing data requirements while maintaining acceptable performance levels. We can express it formally in the following manner: let $N$ be the total number of records and $T(N)$ be the time units required to perform a specific operation. An ODC $\mathcal{C}$ is considered scalable concerning that operation if the condition in Equation \ref{eq:sca} holds.

\begin{equation}
\lim_{{N \to \infty}} \frac{{T(N)}}{N} = 0
\label{eq:sca}
\end{equation}

This condition indicates that the time required for a specific operation per data record decreases as the data volume increases or at least remains bounded. It implies that the data size does not significantly impact performance. It ensures that the ODC can efficiently handle growing data volumes while maintaining acceptable performance levels. This is generally true, except if a single operation requires the processing of several attributes.

\subsection{Timeliness} 
ODCs should ensure the timeliness and freshness of the cataloged data. They should provide mechanisms to monitor and update metadata to reflect changes in data assets, such as new datasets or deprecated entries. Timely updates enable users to rely on the most recent information.

\subsubsection{Timeliness in a Data Catalog}
Timeliness in an ODC ensures that the catalog contains up-to-date information and can be evaluated from two perspectives: data freshness and availability.

\subsubsection{Data Freshness}
Data freshness is the quality dimension that aims to quantify the extent to which the information in the ODC reflects the current state of its underlying data. More formally, let $C(a_k)$ be the timestamp associated with attribute $a_k$, indicating the last update time of that attribute. Given a current timestamp $t$, an ODC $\mathcal{C}$ is considered to have \textit{data freshness} if the Equation \ref{eq:time} holds.

\begin{equation}
\forall a_k \in \mathcal{C}, \exists t : t > C(a_k)
\label{eq:time}
\end{equation}

\subsubsection{Data Availability}
Data availability aims to quantify the catalog's responsiveness. Therefore, let $R(a_k)$ be the response time associated with attribute $a_k$, representing the time required to retrieve that attribute. Data availability is about the extent to which the data might be readily accessible within an acceptable time frame. 

We can formally formulate this notion as follows: an ODC $\mathcal{C}$ is considered to have \textit{data availability} if the Equation \ref{eq:time2} holds.

\begin{equation}
\forall a_k \in \mathcal{C}, R(a_k) \leq T
\label{eq:time2}
\end{equation}

where $T$ is a predefined time threshold. 

\section{Cross-Catalog Quality Dimensions}
Cross-catalog dimensions involve more than one ODC at the same time. This work focuses on ODC compatibility and similarity, which are distinct concepts often used interchangeably in data management. Compatibility refers to the ability of an ODC to seamlessly integrate with various data sources and technologies, enabling adequate replacement. It ensures interoperability between ODCs and different data platforms. On the other hand, similarity aims to identify similarities between ODCs. This involves analyzing data content and structure to uncover relationships and patterns. While compatibility focuses on connectivity and interoperability, similarity emphasizes the analysis of data attributes and properties for knowledge extraction.

\subsection{Compatibility}
The compatibility of two ODCs refers to their respective datasets' integration and complementarity. In order to assess compatibility, several key factors must be considered: the format and structure of both catalogs and common identifiers across datasets that facilitate the merging or replacement process.

Compatibility relies on shared coverage and standardized geographic identifiers or coordinate systems when geographic information is involved. Temporal aspects should also be examined to determine whether the time ranges of both datasets align or complement each other. Finally, data licensing must be carefully evaluated to ensure that both ODCs permit the same actions.

More formally; let $\mathcal{C}_1$ and $\mathcal{C}_2$ be two ODCs. Their compatibility can be defined as shown in Equation \ref{eq:compat}.

\begin{equation}
\text{comp}(\mathcal{C}_1, \mathcal{C}_2) = \frac{|\mathcal{C}_1 \cap \mathcal{C}_2|}{|\mathcal{C}_1|}
\label{eq:compat}
\end{equation}

where $|\mathcal{C}_1|$ is the cardinality of $\mathcal{C}_1$,and $\mathcal{C}_1 \cap \mathcal{C}_2$ is the intersection of $\mathcal{C}_1$ and $\mathcal{C}_2$. Therefore, this dimension estimates the cardinality ratio of the intersection between the two ODCs to the cardinality of $\mathcal{C}_1$ (or $\mathcal{C}_2$). In this way, the compatibility score can range from 0 to 100, where a value of 0 indicates that both catalogs are incompatible since they have no common elements. In contrast, a value of 100 indicates that the catalogs are compatible since all their elements are common.

\subsection{Similarity}
Assessing the similarity between ODCs is crucial in many integration tasks. Similarity assessment enables comparing and aligning ODCs from different sources, facilitating data integration, exchange, and collaboration across systems \cite{key-martinez-eswa2}. Quantifying the degree of similarity between catalogs makes it possible to identify shared attributes and match corresponding items \cite{key-martinez-jfis}. 

Furthermore, similarity assessment can facilitate search and retrieval, enabling users to locate relevant ODCs based on their similarity to a given input \cite{key-martinez-jiis}. This capability enhances data discovery, facilitates reuse, and promotes efficient decision-making processes. Therefore, similarity assessment is essential in promoting data interoperability, enabling effective data integration, and facilitating data utilization across diverse environments \cite{key-martinez-mlwa}.

In this work, we refer to attribute-level similarity, which measures the similarity of the attribute values between corresponding attributes in the two ODCs. The idea is to quantify the resemblance between attribute values based on appropriate similarity measures, such as cosine similarity or Jaccard similarity. Therefore, the attribute-level similarity $\text{Sim}_{\text{attribute}}(\mathcal{C}_1, \mathcal{C}_2)$ can be defined as shown in Equation \ref{eq:sim2}.

\begin{equation}
Sim_{attribute}(\mathcal{C}_1, \mathcal{C}_2) = \frac{1}{K} \sum_{i=1}^{K} Sim_{value}(O_i, O'_i)
\label{eq:sim2}
\end{equation}

where $\text{Sim}_{\text{value}}(O_i, O'_i)$ represents the similarity between the attribute values of the corresponding attributes in objects $O_i$ and $O'_i$.

\section{Non-Core Quality Dimensions}
In the context of an ODC, non-core quality dimensions refer to additional aspects of the datasets that complement the core quality dimensions. While core quality dimensions are essential for evaluating the intrinsic properties of the data (e.g., accuracy, completeness, consistency, etc.), non-core quality dimensions provide supplementary information that enhances the overall understanding and usability of the datasets.

Non-core quality dimensions enrich the dataset's metadata by providing additional context, usage guidelines, and documentation. The idea is to ensure that users understand the dataset well beyond its intrinsic properties. Such context is pivotal in guiding users toward making informed decisions. Incorporating non-core quality dimensions within an ODC promotes transparency, accessibility, and optimal utilization of OD resources. Therefore, we will investigate how these non-core quality dimensions for ODCs can be effectively evaluated in the subsequent subsections.

\subsection{Lineage and Provenance} 
ODCs should incorporate data lineage and provenance information whenever possible. Data lineage tracks the history of data transformations and provides insights into how data has been derived, increasing transparency and aiding data quality assessment. Provenance information captures the source and history of data assets, enhancing credibility.

Therefore, this dimension assesses the lineage and provenance of an ODC. It should iterate through each dataset in the catalog and calculate a lineage score and a provenance score based on the presence of lineage information, ancestors, descendants, provenance information, data sources, and data processing steps. This is just an idea for implementation, but the resulting average lineage and provenance scores enable researchers and practitioners to comprehensively evaluate the historical relationships and origins of the cataloged data.

\subsection{Readability} 
The readability of free-form text in an ODC is crucial for comprehension and data usability. Clear and concise language facilitates efficient data discovery and other potential applications without unnecessary hurdles. Transparent communication of text facilitates responsible data utilization. Additionally, readable catalogs build trust and credibility, encouraging data sharing and stakeholder collaboration.

This dimension aims to assess the readability of an ODC. It should calculate the average readability score by considering the Flesch-Kincaid Grade Level \cite{key-martinez-read} for the title and description of each dataset in the catalog. Using established metrics for readability assessment, such as the Flesch-Kincaid Grade Level, provides a quantitative measure of the catalog's readability, which can help evaluate its accessibility and comprehensibility to a broader audience.

\subsection{Licensing} 
In ODCs, licensing is pivotal in shaping the dissemination and utilization of information. ODCs serve as repositories of diverse datasets, facilitating transparency and collaboration among the data stakeholders. Within this context, implementing suitable licensing mechanisms becomes imperative to ensure equitable access, protect intellectual property rights, and facilitate responsible utilization. This dimension is designed to detect the presence of licensed items within a given ODC. It would also be possible to operate by iteratively examining each item in the catalog, if necessary.

\section{Use cases}
To illustrate how our methods work in practice, we are going to test them in a real environment. For this, we are going to select a set of ODCs of public institutions at the European level and test our quality measures live. 

\subsection{Empirical testing}
We will see how our proposal works with some ODCs offered officially through the European Union's open data portal\footnote{\url{https://data.europa.eu/}}. The objective is to determine how our proposal performs in a real environment and compare these results with the quality rating given by the European Union's open data portal.

To do that, we have chosen several ODCs and will study each of the quality dimensions described earlier in this paper. The criteria for the selection of these ODCs have been that 

\begin{itemize}
	\item they did not belong to any specific country of the European Union but rather to any of its agencies or organizations, 
	\item they were represented using the DCAT standard, and 
	\item they represent a balance combination of catalogs, datasets, and distributions.
\end{itemize}
 
It is necessary to remark that a slight adaptation has had to be made since many ODCs indistinctly combine DCAT tags with Dublin Core \footnote{\url{https://dublincore.org/}} tags, which is a widely used standard in this context.

It should be noted that the experiment was performed on March 21, 2024, so some of the results may not be reproducible. To minimize this risk, we have stored a copy of the ODC available at that date in our repository. However, some aspects, such as the accessibility of the sites listed in the ODC, cannot be reproduced exactly\footnote{The reason is that, for example, a website might not always have continuous availability}. As a result of our filtering process, we have selected the following ODCs for our evaluation:

\begin{itemize}
	\item \textbf{HADEA}: European Health and Digital Executive Agency\footnote{\url{https://data.europa.eu/data/catalogues/hadea}}
	\item \textbf{Euromaps}: Open Maps of Europe\footnote{\url{https://data.europa.eu/data/catalogues/eurogeographics}}
	\item \textbf{Europeana}: Cultural heritage datasets aggregated by Europeana\footnote{\url{https://data.europa.eu/data/catalogues/europeana}}
	\item \textbf{European Union Aviation Safety Agency}: Data regading safety in aviation\footnote{\url{https://data.europa.eu/data/catalogues/easa}}
\end{itemize}

Table \ref{tab:tab1} shows the results obtained for each ODC. We have studied neither similarity nor compatibility since these ODCs belong to fields of knowledge that are too different for such a comparison to make sense. In addition, the last row indicates the average rating the EU gives to each ODC we have considered. This average rating can be \{Excellent, Good, Sufficient\}. The European Union calculates this measure for each dataset, so we show an average for that rating here.

\begin{table}[]
\caption{Summary of the results obtained for ODCs offered through the EU open data portal}
\label{tab:tab1}
\centering
\begin{tabular}{|l|l|l|l|l|}
\hline
    
						& \textbf{HADEA} & \textbf{Euromaps} & \textbf{Europeana} & \textbf{EASA} \\
\hline
Accuracy     &   90.54\% 		& 94.33\% & 89.08\% & 82.97\% \\
Completeness &   88.90\%    & 81.48\%  & 77.78\% & 83.34\% \\
Consistency  &   96.67\%    & 40.00\% &  80.56\% & 72.67\% \\
Scalability  &   Yes    & Yes  &  Yes & Yes \\
Timeliness   &   Yes    & Yes  &  Yes & No \\
Provenance   &   0\%    & 0\%  &  0\% & 0\%\\
Readability  &   17.70    & 15.13  &  10.60 & 7.90 \\
Licensing    &   25\%   &  40\% &  100\% & 100\% \\
Official EU assessment &    Good  &  Good & Good  & Sufficient \\
\hline
\end{tabular}
\end{table}

As can be seen, there is a high degree of correspondence between the EU's rating on its portal and the one obtained automatically and shelled using our approach.

\subsection{Discussion}
We have seen how our framework contributes to metadata data management by providing a practical approach for researchers and practitioners to assess the quality of ODCs. Adopting this framework allows data consumers to establish a robust foundation for data-driven initiatives, improving reliability and enabling trust in the cataloged data.

Several dimensions have been addressed in one way or another in the literature, but the framework presented here does not incorporate them. Below, the reasons are explained:

\begin{enumerate}
	\item Accessibility \cite{key-Lobe}: This is considered a particular case of the notion of accuracy, as a non-functional link can impact the accuracy of the represented information.
	\item Findability \cite{key-wilkinson}: It is seen as a particular case of the notion of completeness, where more reported information increases the chances of finding assets.
	\item Degree of openness \cite{key-Neumaier}: Although crucial for the OD movement, the proposed framework does not consider it a quality-related aspect. For instance, different file formats (.xlsx vs. .csv) do not provide clues about the higher value.
	\item Syntactic inconsistency \cite{key-Debattista}: Parsers automatically detect this type of inconsistency, and the ODC is assumed to be well-formed.
	\item Profile inconsistency \cite{key-kirstein2}: The framework's mission is to remain as generic as possible, and thus, it does not cover situations where specific information needs to impose restrictions on certain values.
\end{enumerate}

However, promising lines exist for future research and development in this domain. Expanding the framework to incorporate additional quality dimensions is one potential direction. Potential novel dimensions such as interpretability or security can lead the framework to provide a more comprehensive quality assessment. 

Another approach for future exploration involves leveraging machine learning to enhance the accuracy of the quality assessment framework. Integrating deep learning and natural language processing techniques can enable the framework to uncover quality patterns and anomalies, entrusting data stakeholders with deeper insights into the cataloged data.

Furthermore, it is crucial to validate the framework on a wide range of domains and use cases. Conducting empirical studies and real-world evaluations across diverse industries and application scenarios will help demonstrate the framework's applicability.

\section{Conclusion}
This work has presented a framework for automatically determining the quality of ODCs. The framework provides a comprehensive approach by addressing an efficient quality assessment in the face of increasing data volumes and diversity. The framework's primary objective is to provide a comprehensive approach that ensures a reliable assessment of ODC quality, given the ever-growing data volumes and the inherent diversity of data sources and formats. Our proposed method facilitates a systematic evaluation across multiple quality dimensions, encompassing core metrics such as accuracy, completeness, consistency, scalability, and timeliness, as well as compatibility and similarity when comparing catalogs, and non-core metrics such as lineage and provenance, readability, and licensing.

Our goal is to provide OD consumers with a sophisticated solution for quality assessment. The framework equips stakeholders with the means to establish sound data management and governance practices. The core idea is to help assess data assets' reliability and utility, promote effective data utilization, and facilitate informed decision-making processes. Our proposed implementations using the DCAT vocabulary showcase the framework practical application framework's practical application, demonstrating its effectiveness in identifying and quantifying quality issues. 

Our research results give us an idea about the significance of automated quality assessment in enhancing the usability and trustworthiness of ODCs, thus facilitating efficient data discovery, integration, and decision support. Therefore, the framework presented in this paper offers data stakeholders a solution for ensuring metadata quality. 

Future research directions may focus on expanding the framework to incorporate additional quality dimensions, exploring advanced machine learning algorithms for quality assessment, and validating the framework on a broader range of domains and use cases. The idea behind continuously striving for improvement and exploring new dimensions is to encourage the research community to drive progress and innovation, ensuring the effectiveness of data-driven initiatives.

\section*{Acknowledgments}
The research reported in this paper has been funded by the Federal Ministry for Climate Action, Environment, Energy, Mobility, Innovation, and Technology (BMK), the Federal Ministry for Digital and Economic Affairs (BMDW), and the State of Upper Austria in the frame of SCCH, a center in the COMET - Competence Centers for Excellent Technologies Programme managed by Austrian Research Promotion Agency FFG.

%

\bibliography{mybib}

\end{document}